\documentclass[prb,aps,twocolumn,floats,showpacs,amsmath,amssymb]{revtex4}
\usepackage{bm,graphicx}
\newcommand{\be}{\begin{equation}}
\newcommand{\ee}{\end{equation}}
\newcommand{\bea}{\begin{eqnarray}}
\newcommand{\eea}{\end{eqnarray}}
\newcommand{\phrl}[1]{Phys.~Rev.~Lett. {\bf #1}}
\newcommand{\phrb}[1]{Phys.~Rev.~B {\bf #1}}

\newcommand{\bib}{\bibitem}

\begin{document}

\title{Orbital Effects of Strong Magnetic Field on a 2-D Holstein Polaron}
\author{S Pradhan$^1$}
\email[E-mail: ]{spradhan@phy.iitkgp.ernet.in}
\author{Monodeep Chakraborty$^1$}
\email[E-mail: ]{bandemataram@gmail.com}
\author{ A. Taraphder$^{1,2}$ }
\email[E-mail: ]{arghya@phy.iitkgp.ernet.in }
\affiliation{ $^1$Department of Physics, Indian Institute of Technology Kharagpur, Kharagpur - 721302, India \\ $^2$Center for Theoretical Studies, Indian Institute of Technology Kharagpur, Kharagpur - 721302, India}

\begin{abstract}
We investigate the orbital effects of strong external magnetic field on the ground state properties of a two dimensional holstein polaron, employing variational approaches based on the exact diagonalization (VAED). From the ground state energy and the wave function we calculate electron-phonon correlation function, the average phonon number and the Drude weight and investigate the evolution of a 2D holstein polaron as a function of the magnetic flux. Although the external magnetic field affects the polaron throughout the parameter regime, we show that the magnetic field has a stronger effect on a loosely bound (spatially extended) polaron. We also find that the magnetic field can be used as a tuning parameter, particularly for weakly coupled polaron, to reduce the spatial extent of a large polaron.

\end{abstract}
\date{\today}
\maketitle

\section{Introduction}
The interplay between electronic and lattice degrees of freedom is central to many areas of condensed matter physics. The holstein model\cite{hol}, which is almost seven decades old, still holds a place of eminence, when it comes to electron-phonon (el-ph) interaction, because of the intricate many-body physics it encompasses within a simple paradigm. On the other hand, modification of electronic band structure on a lattice in the presence of strong magnetic field gives rise to the well-known Hofstadter's butterfly\cite{Hof}, a quintessential example
of fractals in quantum mechanical systems. Although the physics of solids is replete with electron-phonon coupled systems, precious little is known about the effects of strong magnetic field on such a coupled system, where indeed only one of the partners, the electron, feels the effect of the field directly. 

In this paper we study the holstein polaron in a strong magnetic field, which marries the two issues above. In the presence of strong el-ph coupling, the electrons are expected to form  polaronic bound states of varying size, depending on the strength of el-ph interaction. On the other hand, an orbital magnetic field induces a precession of an electron away in an orbit reducing the net mobility, in effect subjecting it further to localization by phonons. It is in this context, the combined effect is an interesting puzzle calling for a resolution. However, the subject of lattice polarons in a strong magnetic field is largely an unexplored area, barring very few work, notably by Mona Berciu\cite{Berciu1}, using momentum-average (MA) approximation. The variational approaches based on the exact diagonalization (VAED) used by Bonca, et al.~\cite{Trug1,Trug2,Trug3,Trug4} and Chakraborty, et al.~\cite{Mono1,Mono2,Mono3,Mono4,Mono5} is one of the most successful numerical methods to study the holstein and extended-holstein type of el-ph systems in dilute regime in all dimensions.

We have generalized this method to deal with cases where the lattice unit cell has more than one equivalent site, i.e., a 
supercell-VAED. This scheme has been very successfully implemented in the present case and can be used to many  important situations where a many-atom unit cell is coupled to
another degree of freedom (like polarons in graphene or a case of partial disorder). We first compare our supercell-VAED zero field results to the benchmark results available in the literature, where we find at least a $8$-digit match. Then we proceed to study the ground state Hofstadter band at different parameters of the holstein model. We first create a variational space by repeated action of the hamiltonian on the initial state and then we adopt a two fold approach: (i) integrate the spectral function obtained from k-space Green's function over the brillouin-zone to get the density of states and (ii) find the ground state energy and wave-function by employing conjugate gradient technique\cite{vishwanath}. 

This paper is organized as follows: in section II, we  discuss the hamiltonian and delineate the basis generation procedures. We then proceed to show our results and compare some of those with existing results in section III. In this section, we also study the evolution of the polaron with magnetic flux in different parameter regimes and try to analyze the interplay between the el-ph interaction and the magnetic field. Conclusion follows in Section IV.

\begin{figure}[t]
\includegraphics[scale=0.33]{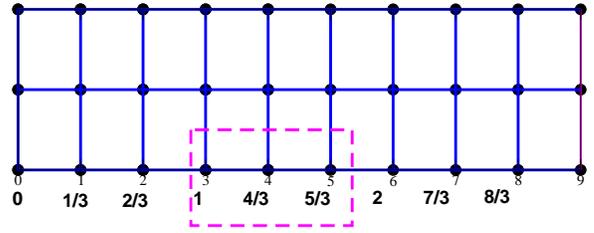}
\caption{\label{f1} (Color online)
A 2-dimensional square lattice in a magnetic field of $\frac{2\pi}{3}$. The figure shows the variation
of magnetic flux along the y-bonds. 
}
\end{figure}

\section{The Model}

In order to study the effect of magnetic field, we employ the simplest and well-studied model for electron-phonon interaction, the holstein model, where a spinless electron is coupled to a dispersionless optical phonon, represented by a local Einstein oscillator. For a spinful model, the magnetic field leads to the Zeeman splitting of up and down-spin polaron bands. In a very high magnetic field and low electron density, one can consider only orbital effects and ignore the Zeeman splitting as the Hilbert space is spin-split into two sectors well separated in energy. One can, therefore, concentrate on the lower of the two for low filling, effectively reducing the model to a spinless one. The hamiltonian is given by,

\bea
H = &&- \sum_{i,j}t_{ij}( c_{i}^{\dag} c_{j} + h.c)
+ \omega \sum_i b_i^{\dag} b_i \nonumber \\
&&- \omega g \sum_{i}c_{i}^{\dagger}c_{i}  (b_{i}^{\dag}
+ b_{i}) 
\eea

\noindent where ${<i,j>}$ are near-neighbour site indices on a square lattice, $c_{i}$$( b_{i})$ are electron (phonon) annhilation operators respectively. The nearest neighbour 
hopping integral in presence of magnetic field is now 
associated with a Peierls phase factor. The choice of a landau gauge $\vec{A}(r)={B(0,ma,0)}$ for a uniform magnetic field $B$  perpendicular to the plane of the lattice leads to the hopping integral $t_{ij} = -t$ along x-direction and $ t_{ij}$=$ -te^{ie/\hbar\int_j^i{A(\vec{r})d\vec{r}}}$=$- t\exp ( \pm 2\pi im\frac{\phi }{{{\phi _0}}})$= $- t\exp ( \pm 2\pi im\frac{p }{{{q}}}) $ along y-direction. Here ${\phi}$=$Ba^{2}$ is the number of flux quanta per plaquette, which is the gain of phase by an electron hopping round a closed path along the plaquette. $\frac{\phi }{{{\phi _0}}}= \frac{p}{q}$ with ${p},\, {q}$ co-prime integers and $\phi_0$ is the Dirac flux quantum. We set the hopping integral ${t}$ to be ${1}$ throughout the numerical calculation and all other parameters are defined in units of ${t}$. Here ${\omega}$ is the oscillator frequency and $g$ is the dimensionless electron-phonon coupling strength. The effect of electron-phonon coupling is expressed in terms of two dimensionless parameters ${\alpha }={\omega/t}$ and $g$.

\begin{figure}[t]
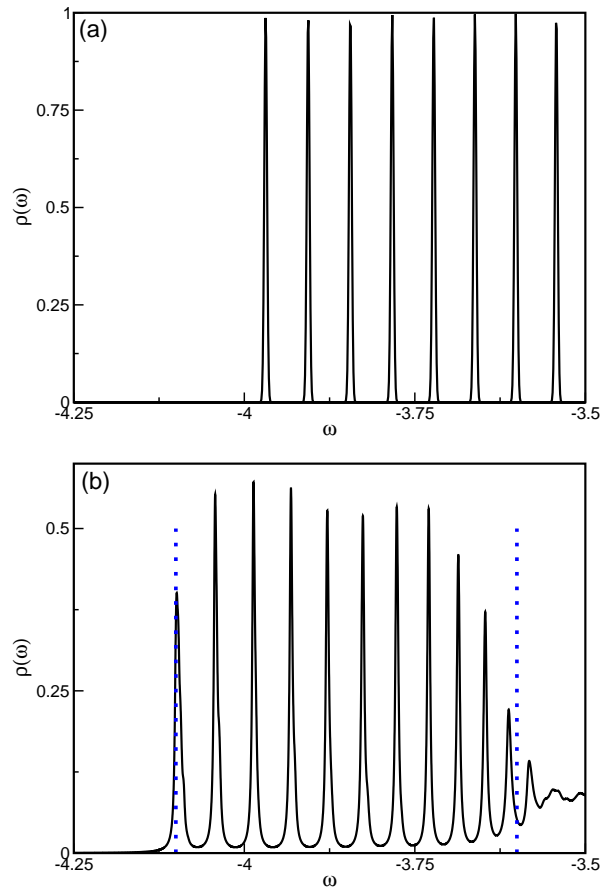

\includegraphics[scale=0.3]{fig2_1.eps}
\vskip 0.3cm
\includegraphics[scale=0.3]{fig2_2.eps}
\caption{\label{f2} (Color online)
The upper panel (a) shows the lower lying landau levels for an electron in a lattice. The lower panel shows the same for a holstein polaron calculated at $\omega$=$0.5$ and $g$=$\sqrt0.4$. The dashed vertical lines mark the ground state energy
at $E_0$ and $E_{0} + \omega$.}
\end{figure}

Once the magnetic field is switched on, we lose lattice periodicity in x-direction. However, since it is only a phase,
repeating at every $2\pi$, periodicity is still retained, albeit, with a changed value. Fig.1, describes the situation 
for a magnetic field $B$=$\frac{2\pi}{3}$. In this case the changed periodicity of the 
lattice is $3$, so instead of the one site unit cell, our magnetic unit cell is basically a $3$-site strip. Therefore to
account for a magnetic field $B$=$\frac{2\pi}{N}$, the magnetic supercell will be a strip of length $N$. The variational 
basis is similar to that of Bonca et. al.\cite{Trug1}, the difference being, now instead of one initial zero phonon state,
we start with ``$N$" zero-phonon states (for $N$ different positions of the magnetic strip) and while checking for the translational symmetry, we shift the supercell. Mathematically, this amounts to putting another index ``$m$", to account for the position-dependent parameter.  In order to get a convergent basis for small polaron ($\omega$=$5.0$ and $g$=$2.0$)
we have appropriated the idea of Lang-firsov transformation while constructing the basis.\cite{Mono3,Mono4,Zhou} This method works very well for the case
of small polarons as has been established by earlier works as it incorporates
into the basis the important states which a small polaron requires.\cite{Mono3,Mono4,Zhou}

We calculate the following quantities of interest: To calculate the correlation between the electron position and phonon distribution (lattice deformation) in the ground state, we define a correlation function as follows:

\begin{equation}
\chi \left( {i - j} \right) = \left\langle {\left. {{\psi _G}} \right|c_i^\dag {c_i}\left( {b_j^\dag  + {b_j}} \right)\left| {{\psi _G}} \right.} \right\rangle 
\label{eq:chi}
\end{equation}

\noindent where ${\psi _G}$ is the ground state wave-function. The total lattice deformation is conserved to $2g$ in the hamiltonian,  from a straightforward sum-rule. The average phonon number is calculated as 

\begin{equation}
{N^{ph}} = \sum\limits_i {\left\langle {\left. {{\psi _G}} \right|\left. {{b_i}^\dag {b_i}} \right|{\psi _G}} \right\rangle } 
\label{eq:phonon_no}
\end{equation}

\noindent The distribution of number of phonons in the vicinity of the electron is given by 

\begin{equation}
\gamma \left( {i - j} \right) = \left\langle {\left. {{\psi _G}} \right|c_i^{{\rm{\dagger}}}{c_i}\left( {b_j^{{\rm{\dagger}}}{b_j}} \right)\left| {{\psi _G}} \right.} \right\rangle 
\label{eq:gamma}
\end{equation}
 The Drude weight ($D_0$)of the ground state of the polaron is obtained by introducing a phase factor to the hopping matrix 
elements ($t \rightarrow t e^{i \eta}$)  and then finding out the response 
 to the electric current as 

\begin{equation}
D_0 =  \frac{\partial^2E_0(\eta)}{\partial \eta^2} |_{\eta=0}
\end{equation}
where $E_0(\eta)$ is the eigen energy of the ground state in
presence of non zero $\eta$.\cite{Mono2} The calculated $D_0$ has been normalized with respect to a free electron
on a square lattice for all the cases.

\section{Results }

We first test the numerical accuracy of our result, by comparing the ground state energy of a holstein polaron at $\omega=2.0$ and $g=1.0$ with the best available results in the literature~\cite{Trug2,Mono4}. The ground state energy for this parameter is $-4.81473577$, obtained from a vartiational basis constructed by operating the hamiltonian $11$ times on the initial states ($N_h$=$11$~\cite{Trug1}), which matches up to $8$ decimal places with the result of Bonca et al.\cite{Trug2}. Then we compare the  density of states (DOS) with the inclusion of el-ph coupling with
the results of Berciu, et al.~\cite{Berciu1}. Fig.2 displays the landau levels for a magnetic flux ($\phi$) of strength $0.005$, for an electron (fig2.(a)) and that of a polaron (fig2.(b)) at $\omega=0.5$ and $g=\sqrt0.4$ (corresponding to $\lambda=0.2$ of Berciu~\cite{Berciu1}). In order to accommodate a flux of value $0.005$, the size of our magnetic supercell (or magnetic strip)had to be $200$. Our results are in excellent agreement with earlier results~\cite{Berciu1} wherever a comparison is possible. The vertical dotted lines of fig 2(b) indicate the ground state energy ($E_0$) and $E_0$+$\omega$ respectively. The distinct polaron landau levels lose their identity beyond $E_0$+$\omega$. We digress a bit and compare the scenario of an electron in a lattice with that of a holstein polaron without bringing the magnetic field into consideration. The electron has only a single band, whereas a holstein polaron has infinite number of bands as the electron is coupled to all possible phononic excitations. However, for holstein polaron we have a gap of $\omega$ at $k$=$0$ in between the ground state and the first excited state, beyond that we have huge number of closely spaced states (depending on the parameter regime) and as we go up in energy they acquire a quasi-continuum nature. These are the states that disturb the sharpness of the landau levels beyond $E_0$+$\omega$ in fig.2(b). 

We now investigate the response of the polaron to magnetic fields. The average phonon number ($N_{ph}$) gives
an idea about the phononic activity of the polaron. Increase in the value of $N_{ph}$ suggests an increase in phononic 
attributes of the polaron. Fig.3(a) shows the variation (relative to zero-field value) of $N_{ph}$ at three different regimes. At $\omega=1.0$ and $g=0.05$, we have a quasi-free electron at $B=0$, and the phonon activity clearly increases with the increase in field values, achieving a maximum in the vicinity of $\phi=0.375$. Though the trend is similar at $\omega$=$2.0$ and $g$=$1.00$, the relative variation is much less pronounced. At $\omega=5.0$ and $g=2.0$, we are deep in the anti-adiabatic limit ($\omega \gg t$), polarons of small spatial extent hardly respond to the variation in magnetic field. Fig.3(b) shows the relative variation (w.r.t. the zero field value) of ground state energy with field for the same set of $el-ph$ parameters. The pattern in ground state energy variation is same for all the three parameter regimes and one can recognize them as ground state Hofstadter band. The relative change is again maximum for a weakly bound polaron and minimum for the polaron in the anti-adiabatic limit; the signature of magnetic field shows up in ground state energy pattern for all regimes. The variation in average phonon number as well as ground state energy with applied magnetic flux follow the lowest branch of the Hofstadter butterfly. Both fig.(3(a) and 3(b)) are symmetric about $\phi$=$0.5$, a typical Hofstadter characteristics.  The maximum size of the magnetic strip used for this calculation was $64$.


\begin{figure}[h]
\includegraphics[scale=0.35]{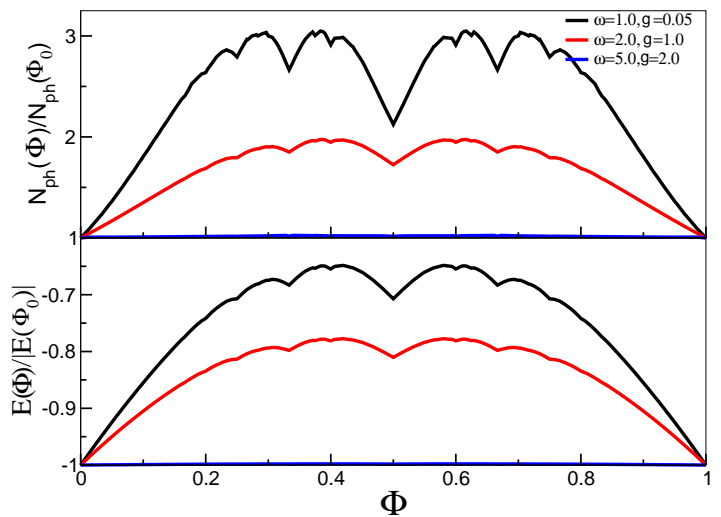}
\caption{\label{f3} (Color online)
The upper panel shows the average phonon number $N_{ph}$ as a function of magnetic field for three different sets of $el-ph$ coupling; $N_{ph}$ are normalized by their respective zero field values. The lower panel shows the ground state energy $E_0$ (normalized by their respective zero field values) as a function of magnetic field for three sets of $el-ph$ coupling.}
\end{figure}

A thorough study of the electron-lattice correlation functions throws some light on the mechanism of increased phononic activity for a weakly bound polaron and a much smaller effect in the anti-adiabatic regime.  A strip size of $16$ was used to calculate this correlation function. Fig.4 shows the $\chi(x,y)$ for four different values of $\phi$ ($\phi$=$0.0, 0.1875, 0.375$ and $0.5$) for a polaron at $\omega$=$1.0$ and $g$=$0.05$.  The displayed $\chi(x,y)$ has been divided by the total deformation $2g$ for convenience. Though the total lattice distortion of the polaron always adds up to $2g$, we can see that with increase in $\chi$ locally, the local distortion assumes a higher value for $\chi(x,y)=\chi(0,0)$ and its vicinity than farther outside, i.e., we a have a relatively tightly bound polaron (as compared to the $\phi$=$0$ polaron). The $\chi(0,0)$ for $\phi=0$ is $0.025$ and reaches a maximum at $\phi=0.375$ to $0.043$. A similar trend is observed for $\gamma(x,y)$ as well. 

\begin{figure}[t]
\includegraphics[scale=0.7]{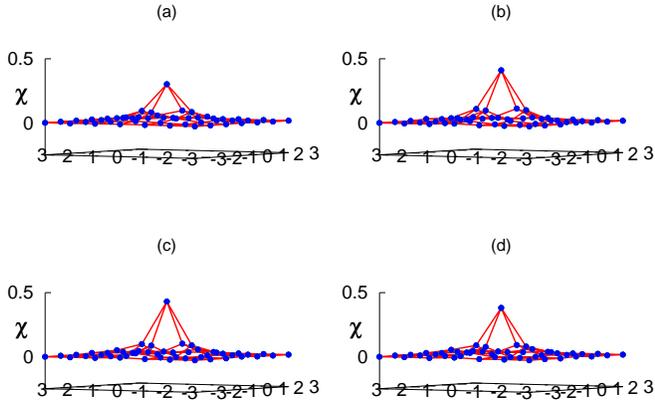}
\caption{\label{f4} (Color online)
The electron-lattice correlation function $\chi(x-y)$ measures the lattice distortion, shown at four different magnetic flux values. The $el-ph$ parameter for the polaron is $\omega$=$1.0$ and $g$=$0.05$.}
\end{figure}
Fig.5 shows the electron-lattice correlation at same four values of $\phi$ ($\phi$=0.0, 0.1875, 0.375 and 0.5) for a polaron in the anti-adiabatic regime ($\omega$=$5.0$ and $g$=$2.0$). An extremely small polaron results in this regime and as it is quite evident, almost the entire distortion is rooted at $(x,y)$=$(0,0)$; there is no noticeable change with magnetic field. The $\chi(0,0)$ for $\phi=0$ is $3.90$ and reaches a maximum at $\phi=0.375$ to $3.943$, a very small change. A study of fig.4 and fig.5 clearly shows that the magnetic field brings about a prominent change in a weakly bound polaron compared to a polaron that is tightly bound to the lattice. The magnetic field tends to shrink the distortion towards its centre for a weakly bound polaron, whereas it can hardly affect a strongly bound one. Clearly a loosely bound polaron has a much larger orbit and therefore has a stronger effect due to the orbital magnetic field. Hence, we see in fig.4, the relative increase in $\chi(0,0)$, though the sum total of distortion was limited to $2g$ as expected.
 
\begin{figure}[t]
\includegraphics[scale=0.7]{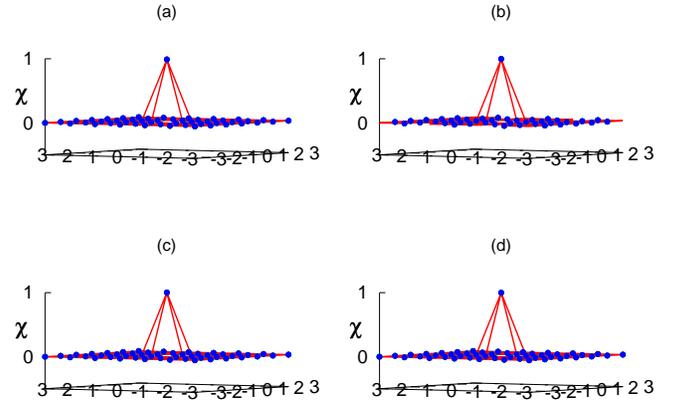}
\caption{\label{f5} (Color online)
The electron-lattice correlation function $\chi(x-y)$, at four different magnetic fields. The respective parameters are $\omega$=$5.0$ and $g$=$2.0$.}
\end{figure}

 The Drude weight ($D_0$) gives the measure of coherence and is an important correlation function to study the nature of 
electronic conductivity of a system. There have been a number of earlier work suggesting magnetic field induced 
metal-insulator transition.\cite{Jin} Fig.6 shows the calculated $D_0$ as a function of field for the three $el-ph$ regimes. 
The size of the magnetic strip used for this calculation was $16$. At $\omega$=$1.0$ and $g$=$0.05$ we have a quasi-free electron very weakly tied to the lattice, with a large spatial extent. Consequently, at $\phi$=$0, \, D_0 \approx 1.0$. However for very small magnetic flux ($\phi$=$\frac{1}{16}$) it almost drops to $0$, suggesting a complete loss of coherent hopping. The inset 
(a), (b) and (c) show the ground state band for five different flux values ($\phi$=$0.0$, $\frac{1}{16}$, $\frac{1}{4}$,
$\frac{3}{8}$ and $\frac{1}{2}$) for three different el-ph regimes. The inset (a), (b) and (c)  at $\phi$=$\frac{1}{16}$ show a completely flat band (within the numerical accuracy of
our calculation) and $\frac{dE_{k}}{dk}$=$0$ throughout the brillouin zone which arises from local quantum 
interferences.\cite{Wolfgang}
 The electron, while moving through the lattice, will see a maximum variation in flux for small values of field and loses its coherence, however the ground state
band is not completely flat,  for $\phi = \frac{1}{4},\, \frac{3}{8},\, \frac{1}{2}$ etc., we can find peaks in  $D_0$. 
  The $D_0$ peaks for $\omega$=$2.0$ and $g$=$1.0$, have slightly reduced values as we are in the intermediate el-ph coupling regime and we no longer have the quasi-free electron. For $\omega$=$5.0$ and $g$=$2.0$, anti-adiabatic strong coupling regime, the exponential suppression of the $D_0$ is seen, however the change in magnetic flux still has an effect, which can be seen in the inset (d) of fig. 6 and  local quantum interferences still has its effect at small magnetic flux ($\phi$=$\frac{1}{16}$), which leads to localization and consequent vanishing of $D_0$. We see a magnetic field induced
coherent to incoherent transition at $\phi$=$\frac{1}{16}$ across the el-ph
coupling regime. 
\begin{figure}[t]
\vskip 0.6cm
\includegraphics[scale=0.34]{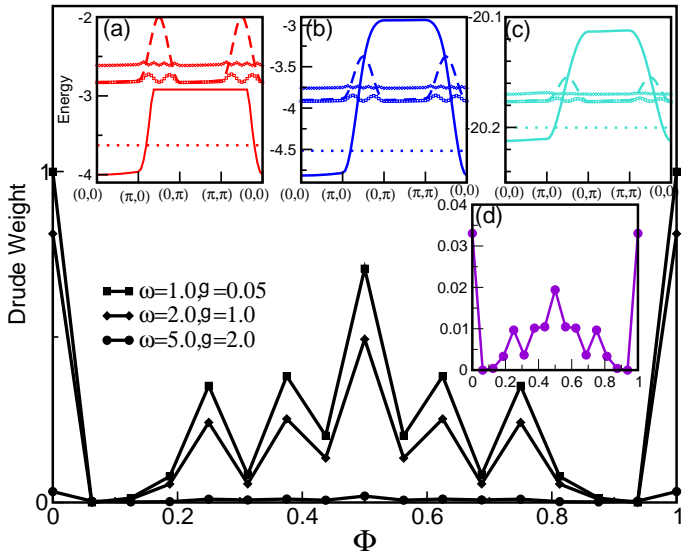}
\caption{\label{f5} (Color online)
Drude weight $D_0$ (normalized by the free electron $D_0$) versus magnetic flux. The inset (a), (b) and (c) shows the ground state band at $\phi$=$0.0$ (solid line),
$\phi$=$\frac{1}{16}$ (dotted line), $\phi$=$\frac{1}{4}$ (open square), $\phi$=$\frac{3}{8}$ (open diamond) and
$\phi$=$\frac{1}{2}$ (dashed line) for ($\omega$=$1.0$, $g$=$0.05$), ($\omega$=$2.0$, $g$=$1.0$) and
($\omega$=$5.0$, $g$=$2.0$) respectively.  Inset (d) shows  $D_0$ for  $\omega=5.0$ and $g=2.0$ separately. }
\end{figure}
\section{Summary and Conclusion}
We have developed a numerical scheme based on VAED to study $el-ph$ interaction in presence of strong magnetic field. Our results are in excellent agreement with earlier results wherever a comparison is possible. The external magnetic field changes the orbit of motion of the electron and the effect is more in case of loosely bound polaron already spread out in real space. With the inclusion of magnetic field, we see that the spatial extent of the polaron decreases resulting in a polaron with a tighter binding. Our method can be useful to many important cases with a more complex unit cell.

\section{Acknowledgments}
 We gratefully acknowledge Narayan Mohanta and Sugata Pratik Khastgir for useful and stimulating discussion. We acknowledge the use of the computing facility from DST-FIST (phase-II) Project installed in the Department of Physics, IIT Kharagpur, India.

\end{document}